\documentclass[sigconf]{acmart}
\usepackage[utf8]{inputenc}
\usepackage{graphicx}
\usepackage{textcomp}
\usepackage{comment}
 \usepackage{url}  
 \usepackage{multirow}
 \graphicspath{ {images/} } 
\usepackage{balance}

\newcommand{\xAppName}{SLNET} 
\newcommand{\GH}{GitHub}
\newcommand{\SL}{Simulink}

\newcommand{\xTable}[1]{Table~\ref{#1}}
\newcommand{\xFigure}[1]{Figure~\ref{#1}}

\newcommand{\CorpusCurated}{SLC0}

\newcommand{\totalGithubproject}{1,284}
\newcommand{\totalGithubprojectwithLicense}{232}
\newcommand{\totalGithubprojectinSLNetLegalwithduplicatesanddummy}{231} 
\newcommand{\totalGithubprojectinSLNetLegal}{225}

\newcommand{\totalmodelFilesinGithubinSLNetLegal}{2,088}  

\newcommand{\totalMATCproject}{2,941}
\newcommand{\totalMATCprojectwithLicense}{2,746}
\newcommand{\totalMATCprojectinSLNetLegalwithduplicatesanddummy}{2,728}
\newcommand{\totalMATCprojectinSLNetLegal}{2,612}

\newcommand{\totalmodelFilesinMATCSLNetLegal}{7,029}

\newcommand{\SLNETallTotalProject}{4,225}
\newcommand{\SLNETlicenseTotalProject}{2,978}
\newcommand{\SLNETlegalTotalProjectswithduplicatesanddummy}{2,959}
\newcommand{\SLNETlegalTotalProject}{2,837}

\newcommand{\SLNETlegalTotalModel}{9,117}

\newcommand{\SLNETnoMetricModels}{88}

\copyrightyear{2022}
\acmYear{2022}
\setcopyright{rightsretained}
\acmConference[MSR '22]{19th International Conference on Mining Software Repositories}{May 23--24, 2022}{Pittsburgh, PA, USA}
\acmBooktitle{19th International Conference on Mining Software Repositories (MSR '22), May 23--24, 2022, Pittsburgh, PA, USA}
\acmDOI{10.1145/3524842.3528001}
\acmISBN{978-1-4503-9303-4/22/05}

\title{\xAppName{}: A Redistributable Corpus of 3rd-party Simulink Models}

\begin{document}

\author{Sohil Lal Shrestha}
 \affiliation{%
   \department{Computer Science \& Eng. Dep.}
   \institution{University of Texas at Arlington}
   \streetaddress{Box 19015}
   \city{Arlington}
   \state{Texas}
   \postcode{76019}
   \country{USA}
 }
 \author{Shafiul Azam Chowdhury}
 \affiliation{%
   \department{Computer Science \& Eng. Dep.}
   \institution{University of Texas at Arlington}
   \streetaddress{Box 19015}
   \city{Arlington}
   \state{Texas}
   \postcode{76019}
   \country{USA}
 }
 \author{Christoph Csallner}
 \affiliation{%
   \department{Computer Science \& Eng. Dep.}
   \institution{University of Texas at Arlington}
   \streetaddress{Box 19015}
   \city{Arlington}
   \state{Texas}
   \postcode{76019}
   \country{USA}
   } 
   
   \begin{abstract}
    MATLAB/\SL{} is widely used for model-based design. Engineers create \SL{} models and compile them to embedded code, often to control safety-critical cyber-physical systems in automotive, aerospace, and healthcare applications. Despite \SL{}'s importance, there are few large-scale empirical \SL{} studies, perhaps because there is no large readily available corpus of third-party open-source \SL{} models. To enable empirical \SL{} studies, this paper introduces \xAppName{}, the largest corpus of freely available third-party \SL{} models. \xAppName{} has several advantages over earlier collections. Specifically, \xAppName{} is 8~times larger than the largest previous corpus of \SL{} models, includes fine-grained metadata, is constructed automatically, is self-contained, and allows redistribution. \xAppName{} is available under permissive open-source licenses and contains its collection and analysis tools.  
\end{abstract}


\maketitle

\section{Introduction}

Currently there is no collection of Simulink models that is commonly used in empirical studies. Though there have been previous model collections, they lack fine-grained meta-information, are not self-contained, and are not redistributable due to restrictive or missing licenses---making them hard or impossible to use for most empirical researchers. Given the lack of such a collection, the few existing empirical studies of Simulink models have been limited to proprietary models or a small number of public models~\cite{simulink:Stephan,simulink:evol_Stephan,chowdhury2018icse}.

Deepening our understanding of Simulink models and modeling practices is important, as Simulink is a de-facto standard tool in several safety-critical industries such as automotive, aerospace, healthcare, and industrial automation---for system modeling and analysis, compiling models to code, and deploying code to embedded hardware~\cite{CPS:WolfF15,CPS:Zheng}. Having a large corpus of third-party Simulink models may make it easier for engineers and researchers to produce, reproduce, and validate empirical results about Simulink models, modeling practices, and tools that operate on such models.

The most closely related previous work has studied an initial collection of 391~third-party Simulink models~\cite{chowdhury2018icse} and later extended it to a curated corpus (``\CorpusCurated{}'') of some 1k third-party Simulink models~\cite{Chowdhury18Curated}. Boll et al.~\cite{simulink:corpus:boll2021characteristics} collected an updated version of \CorpusCurated{} and assessed the corpus's suitability for empirical research. While pioneering larger studies and validating that models from such a corpus can be similar to industrial models, these collections consisted of a list of URLs to non-permanent resources~\cite{chowdhury2018icse} and contained models with unclear license information~\cite{Chowdhury18Curated}. These collections were largely manual, which lead to inconsistencies (empty projects, duplicate projects, and missing metadata), relatively modest collection size, and may yield unintended human errors and bias.

To address these limitations, \xAppName{} automates corpus construction and analysis, including data acquisition, cleaning (except for the rarely required manual review of a new license type), metric computation, and packaging. \xAppName{} thereby automatically mines and analyses Simulink models from the two most popular repositories for sharing Simulink models, yielding a collection of thousands of models that is fully self-contained and allows redistribution.

To allow fine-grained selection of Simulink models and projects, \xAppName{} computes several project-level and model-level metrics~\cite{simulink:corpus:boll2021characteristics} and exposes them in a SQL database. \xAppName{} similarly identifies and labels libraries and models that are test harnesses~\cite{simulink:testharness:Simevo}.
To summarize, this paper makes the following major contributions.
\begin{itemize} 
    \item \xAppName{} is 8~times larger than the prior largest known corpus of third-party \SL{} models. \xAppName{} also adds fine-grained metrics,  being self-contained and redistributable.

    \item \xAppName{}~\cite{dataset:slnet} and its tools~\cite{tool:SLnet-Miner,tool:SlnetMetrics} are available under permissive open-source licenses (CC~BY and BSD 3-clause). 
\end{itemize}

\section{Background on Simulink}

\begin{figure}[h]
\centering
 \includegraphics[width=0.85\linewidth]{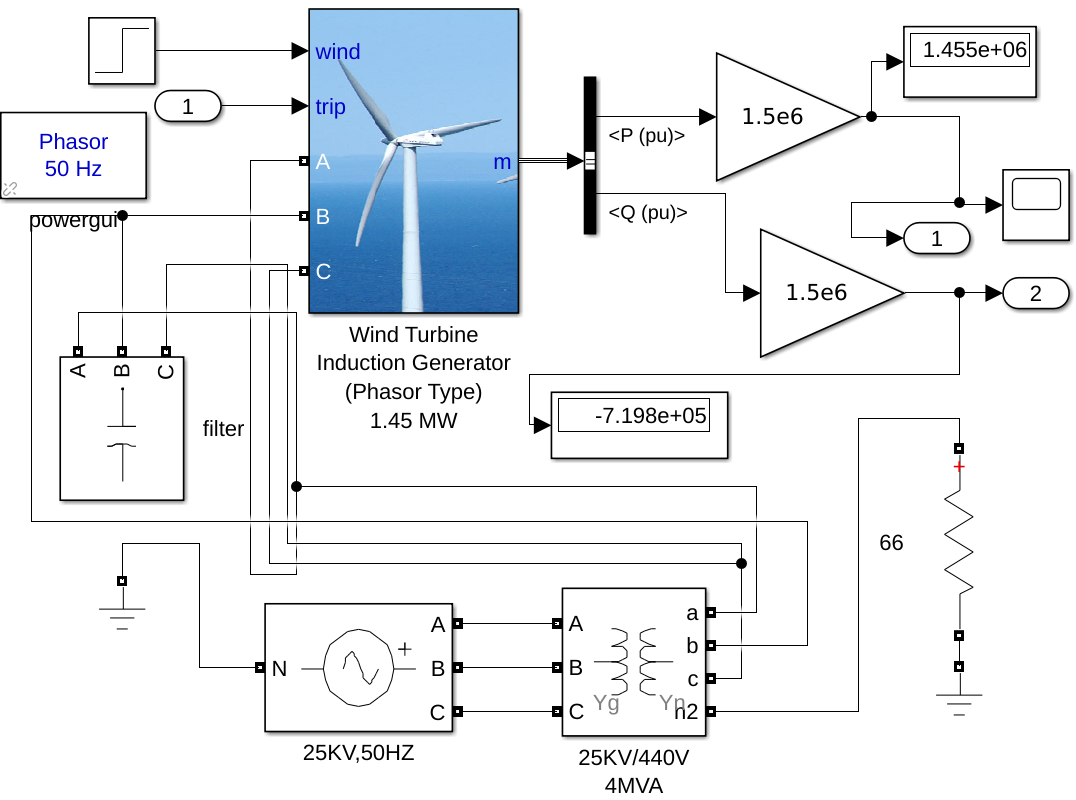}
  \caption{Sample \xAppName{} Simulink model of a 1.5MW wind generation plant~\cite{others:modelfromSLnet} with 18 blocks and 23 connections.}
  \label{fig:toySimulinkModel}
  \centering
\end{figure}

\SL{}~\cite{doc:simulink} is a widely used commercial tool-chain for model-based design~\cite{CPS:WolfF15,CPS:Zheng}\footnote{Searching for ``Simulink'' jobs on LinkedIn in the US currently yields over 5k job postings: \url{https://www.linkedin.com/jobs/search/?keywords="simulink"&location=US}}.
Engineers typically design a cyber-physical system (CPS) model in \SL{}'s graphical modeling environment. A \SL{} model such as \xFigure{fig:toySimulinkModel} is a \emph{block diagram}, where each block represents equations or modeling components. Depending on the block type, each block can accept input (via input ports), perform some operation on its inputs, and produce output (via output ports), which then can optionally be forwarded to other blocks via explicit or implicit connection lines (aka signal lines). \SL{} users can add blocks from various built-in \textit{libraries} and toolboxes~\cite{doc:blocklib}, and can also define \textit{custom blocks} in ``native'' code (e.g., in C) using the S-function interface.

To deal with model size, users can create hierarchical models, by (recursively) grouping blocks in (a) a \emph{Subsystem} or (b) in a separate model via \emph{Model Reference}.
\SL{} does not permit a cyclic model hierarchy, but there may be block connection (aka data dependence) cycles, including algebraic loops\footnote{\url{https://www.mathworks.com/help/simulink/ug/algebraic-loops.html}}. 

As a first step, \emph{compiling} translates the model into a toolchain-internal representation. When \emph{simulating} the compiled model, the toolchain computes the output of each block at successive time steps over a specified time range using pre-configured numerical solvers. \textit{Fixed-step} solvers solve the model at fixed time intervals whereas \textit{variable-step} solvers automatically adjust the time intervals at which the model is solved.
Simulink may reject a model if it cannot numerically solve an algebraic loop. Simulink offers different simulation modes, i.e., \textit{Normal} mode ``only'' simulates blocks, \textit{Accelerator} speeds up simulation by emitting native code, and \textit{Rapid Accelerator} produces a standalone executable\footnote{Simulink's embedded code generation workflow for deployment on target platforms is distinct from these simulation modes.}.

\section{\xAppName{} Design \& Construction}

\begin{figure}[h!tb]
\centering
\includegraphics[trim = .2in .32in .2in .2in, clip, width=.65\linewidth]{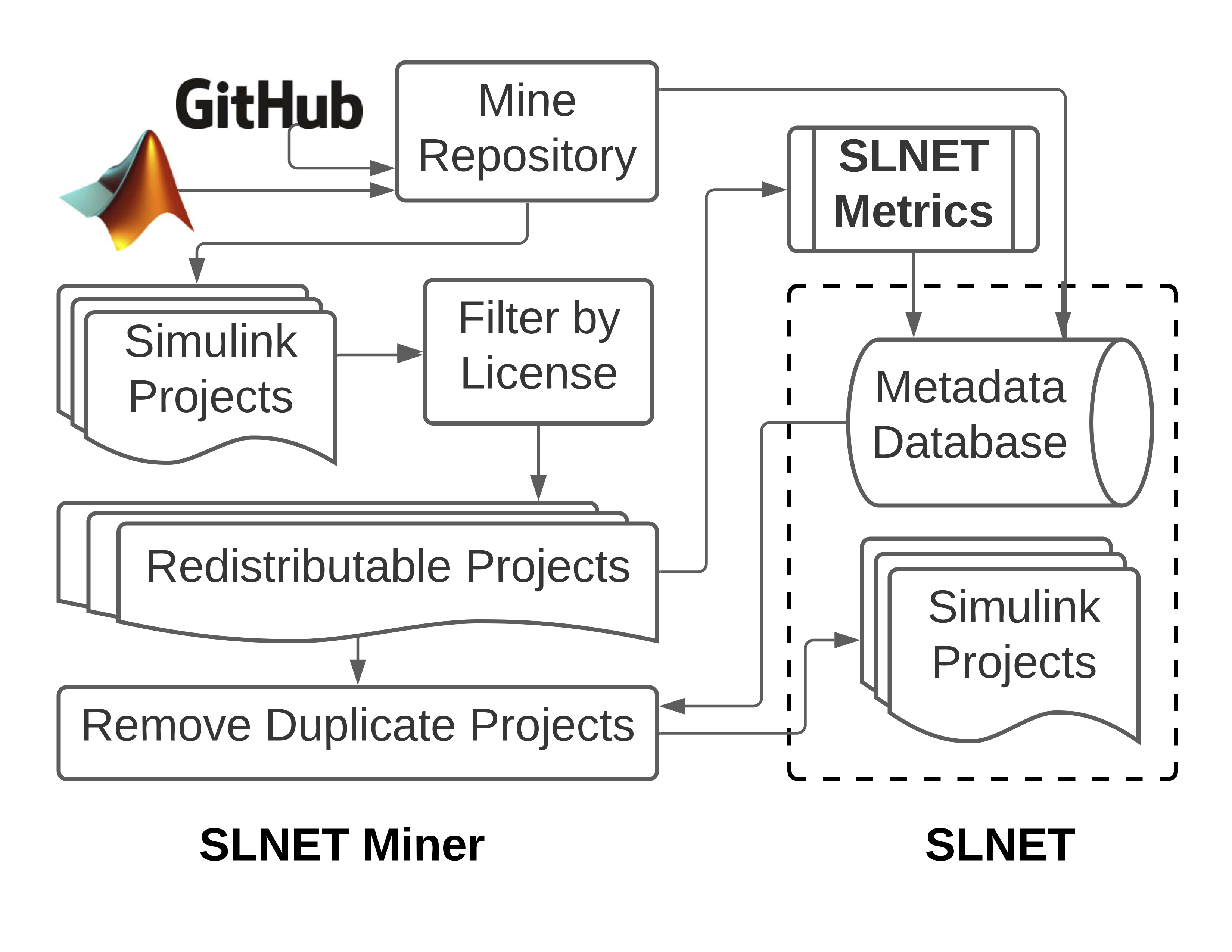}
  \caption{Overview: SLNET-Miner collects files and data, 
  removes empty and duplicated projects or those without appropriate license. SLNET-Metrics extracts
  model metrics.}
  \label{fig:overviewSLNET}
  \centering
\end{figure} 

\xAppName{} is not a superset of earlier \SL{} corpora~\cite{Chowdhury18Curated,simulink:corpus:boll2021characteristics} as earlier corpora were neither self-contained nor redistributable. Figure~\ref{fig:overviewSLNET} gives an overview of \xAppName{}'s construction. We built \xAppName{} from models shared in \GH{}~\cite{repo:github} and MATLAB Central~\cite{repo:matc}. 
Due to time limitations we do not collect \SL{} models from smaller repositories such as GitLab~\cite{repo:gitlab} and  SourceForge~\cite{repo:sourceforge}. Before removing projects that are empty, duplicate, or have an unclear license, a quick search for ``Simulink'' yields some 60 GitLab and some 70 SourceForge projects.

While \GH{} offers commit-level version control, MATLAB Central ``only'' serves project releases. To limit \xAppName{}'s size and due to the different versioning (git commits vs. project releases), in February 2020 we ``only'' collected \SL{} project snapshots (i.e., all current project files plus project metadata).

\GH{} provides a REST API to discover projects and extract them with their metadata. \xAppName{}-Miner queries the \GH{} API (via PyGithub~\cite{library:pygithub}) with the keyword ``Simulink''. Unlike previous work~\cite{chowdhury2018icse,Chowdhury18Curated}, we used keyword search and not file extension search, as file extension search is typically intended to search within a given \GH{} repository and using file extension search in \GH{}'s search page produced many false positives.

The \GH{} API expose 23 types of project-level information~\cite{doc:githubapi}, of which \xAppName{} retains 20. The other 3 are redundant (full project name) or API-internal (API query relevance score and node id). From the API we also obtain each project's topics (user-created labels and tags). From the downloaded project files, we extracted the list of \SL{} model files plus the project's license.

As MATLAB Central ``only'' offers an RSS feed~\cite{repo:rssfeed} for its file exchange platform, we filter the search result feed by \SL{} models and then parse the feed to collect each project's download URL plus 14 other types of project metadata. Since from the RSS feed we could not construct the download URL for all projects, we extracted \totalMATCproject{} of the 3,110 available projects.

\subsection{Data Cleaning \& Storage: ZIP + SQLite}
\label{sec:datacleaning}

We remove projects without \SL{} models (i.e., file extensions \emph{slx} or \emph{mdl}) and projects we know to contain synthetic models (i.e., model generators~\cite{chowdhury2018icse,Chowdhury20SLEMI}). We heuristically search for other model generators (via terms ``automat'', ``random'', ``fuzz'', and ``generate'') in project titles, project descriptions, and project tags, which yielded 530 projects (e.g., on fuzzy logic). As we did not find evidence that these projects generate models we kept them in \xAppName{}.

\begin{table}[h!t]
\centering
\caption{Data cleaning: 
Real~=~has 1+ models (likely non-synthetic);
License~=~has a license;
\xAppName{}+D~=~license allows redistribution;
\xAppName{}~=~has a model with 1+ blocks after removing potential duplicate projects;
Model counts here include 1,130 library and 9 test harness models.
}
\begin{tabular}{lrrrr|r} 
 \hline
 & \multicolumn{4}{c|}{Projects} & Models\\
 & Real & License & \xAppName{}+D & \xAppName{} & \xAppName{}\\
 \hline
GitHub & \totalGithubproject{} & \totalGithubprojectwithLicense{} & \totalGithubprojectinSLNetLegalwithduplicatesanddummy{} &\totalGithubprojectinSLNetLegal{} & \totalmodelFilesinGithubinSLNetLegal{}\\
MATLAB Cl  & \totalMATCproject{} & \totalMATCprojectwithLicense{}&\totalMATCprojectinSLNetLegalwithduplicatesanddummy{}&\totalMATCprojectinSLNetLegal{} &\totalmodelFilesinMATCSLNetLegal{}\\ \hline
Total & \SLNETallTotalProject{} & \SLNETlicenseTotalProject{} &\SLNETlegalTotalProjectswithduplicatesanddummy{}& \SLNETlegalTotalProject{} & \SLNETlegalTotalModel{} \\
 \hline
\end{tabular}
\label{table:projectcompare}
\end{table}

We then remove projects without a license or whose license does not allow redistribution. GitHub has a structured way for authors to set a license, which GitHub converts to a  file (and exposes via an API). We manually reviewed the remaining 50 projects' licenses (where GitHub did not understand the author's license or for MATLAB Central projects without a BSD license).

We heuristically remove potentially duplicate projects. We consider project A a duplicate of B if (1) A and B contain the same number of Simulink model files and (2) there is a bijective mapping between models in A and B based on our Section~\ref{sec:metriccal} model metrics (excluding compile time). If A and B are from the same data source (\GH{} or MATLAB Central), we keep the first-created one in \xAppName{}. Otherwise, we keep the one from GitHub, as it offers more fine-grained meta-data. Finally, we remove dummy projects (projects whose Simulink models all have zero blocks). 

\xTable{table:projectcompare} summarizes data cleaning. After removing model generators we downloaded \SLNETallTotalProject{} projects with at least one \SL{} model, of which \SLNETlicenseTotalProject{} had a license, of which \SLNETlegalTotalProjectswithduplicatesanddummy{} allowed redistribution. Removing 112 potentially duplicate plus 10 dummy projects yielded \SLNETlegalTotalProject{} projects and their \SLNETlegalTotalModel{} \SL{} models in \xAppName{}.

\begin{figure}
    \centering
    \includegraphics[width=0.95\linewidth]{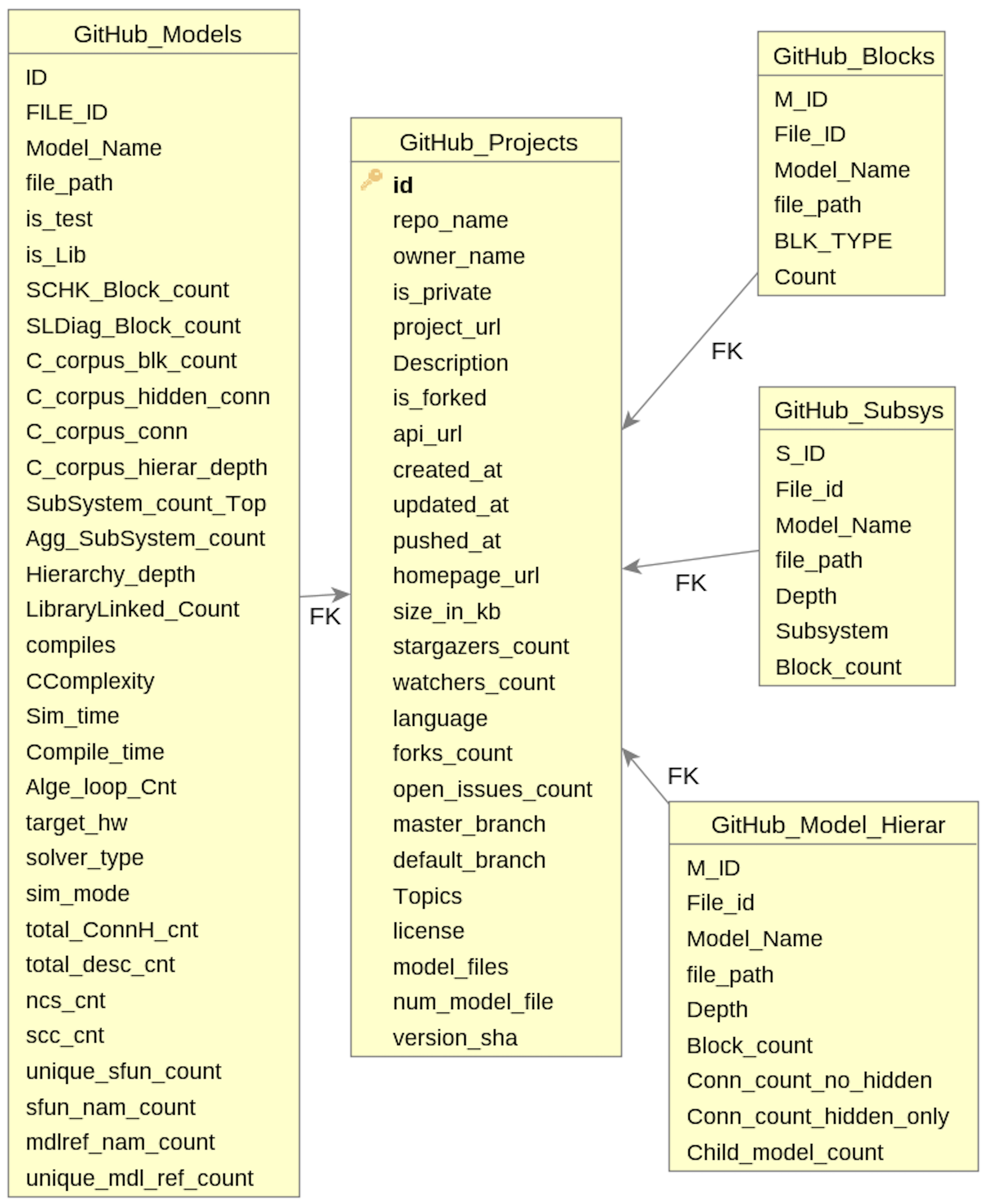}
    \caption{\xAppName{} database schema (GitHub portion). The MATLAB Central portion only differs in its \_Projects table~\cite{dataset:slnet}.}
    \label{fig:SLNET-ER}
\end{figure}

\xAppName{} is on Zenodo (a second archive contains the 112 duplicate projects)~\cite{dataset:slnet}. Each project has a snapshot of its files in a ZIP archive in either the GitHub or MATLAB Central directory. Each project is named \textbf{ID.zip}, where ID is an identifier defined by \GH{} or MATLAB Central. \xAppName{} includes the \xFigure{fig:SLNET-ER} SQLite\footnote{SQLite is widely used, free, self-contained, server-less, zero-configuration, backwards compatible, and cross-platform: \url{https://www.sqlite.org/index.html}} database. It contains project-level information (license type, etc.) from the source repositories and the model metrics our tools extracted. Users can thus select models and projects from \xAppName{} via SQL queries.

\subsection{Project \& Model Metrics}
\label{sec:metriccal}

\begin{table}[h!]
    \centering
    \caption{\xAppName{}'s project engagement distributions are long-tailed as in other studies of open-source projects~\cite{Lima14Coding,empirical:githubtrends,Ma16What,github:perilsofmininggithub}.}
    \begin{tabular}{llcrrcr}
    \hline
   & Metadata & Min & Max & Avg & Med. & SD \\
    \hline
\multirow{3}{*}{GitHub}
& Stargazers	&0	&128	&3.5	&0	&12.1\\
& Forks	&0	&122	&2.8	&0	&10.7\\
& Open Issues	&0	&82	&1.2	&0	&6.5\\
\hline
\multirow{3}{*}{MATLAB Cl}
& Comments	&0	&218	&3.5	&1	&12.3\\
& Ratings	&0	&108	&2.9	&1	&6.8\\
& Avg. Rating	&0	&5	&2.5	&3	&2.2\\
\hline
    \end{tabular}
    \label{tab:github_metadata}
\end{table}

To get an insight into the projects' domain and popularity we first searched the user-generated project tags (i.e., \GH{} ``topics'' and MATLAB Central ``categories'') for common domains (i.e., the \SL{} project domains identified by Boll et al.\cite{simulink:corpus:boll2021characteristics}), yielding
Electronics (983), 
Automotive (64),
Communications (61),
Robotics (52),
Energy (48),
Aerospace (47),
Biotech (20), and
Medicine (2).
\xTable{tab:github_metadata} shows data often used as proxies for project popularity or engagement (e.g., people who have star-ed or forked a \GH{} project or provided a 1--5 star rating for a MATLAB Central project). For example, a \xAppName{} \GH{} project has on average 2.8~forks.

\begin{table*}[h!t]
    \centering
    \begin{tabular}{l|rr|rr|rr|rr|rr|rrrr}
  \hline
  \multirow{2}{*}{} 
      & 
      \multicolumn{2}{c}{Models} &
      \multicolumn{2}{|c}{Hierarchical} &
      \multicolumn{2}{|c}{Blocks} &
      \multicolumn{2}{|c}{Connections} &
      \multicolumn{2}{|c}{Solver Step} &  
      \multicolumn{4}{|c}{Simulation Mode}  \\  
         \multicolumn{1}{l}{Source}  &
         \multicolumn{1}{|c}{M} & 
         \multicolumn{1}{c}{Mc} & 
         \multicolumn{1}{|c}{Mh} & 
         \multicolumn{1}{c}{Mh\textsuperscript{t0}} & 
         \multicolumn{1}{|c}{B} & 
         \multicolumn{1}{c}{B\textsuperscript{t0}} & 
         \multicolumn{1}{|c}{C} & 
         \multicolumn{1}{c|}{C\textsuperscript{t0}} &
          Fixed & Var & Nor & Ext & PIL & Ac \\
          \hline
          GitHub & 1,639 & 541 & 878 & 1,304 & 190,321 & 414,241 & 188,285 & 395,725 & 860 & 762 & 1,501 & 103 & 2 & 14   \\
        MATLAB Cl & 6,251 & 3,636 & 3,893 & 5,566 & 838,956 & 3,197,221 & 915,975 & 3,084,605 & 1,757 & 4,493 & 5,984 & 186 & 2 & 76  \\
        
        \hline
         Total  & 7,890 & 4,177 & 4,771 & 6,870 & 1,029,277 & 3,611,462 & 1,104,260 & 3,480,330 & 2,617 & 5,255 & 7,485 & 289 & 4 & 90  \\ \hline
    \end{tabular}
    \caption{\xAppName{}'s model metrics after removing library \& test harness models;
M~=~models;
Mc~=~models we could readily compile; 
Mh~=~hierarchical models (readily compilable and otherwise);
C~=~non-hidden connections;
\textsuperscript{t0}~=~via \CorpusCurated{}'s metric tool;
Var~=~variable;
Nor~=~normal;
Ext~=~external;
PIL~=~processor in the loop;
Ac~=~accelerator.
For 18 models the API did not indicate simulation mode or solver type. The remaining 4 models are configured for Rapid Accelerator simulation mode.}
    \label{tab:replication}
\end{table*}

To extract commonly used model metrics (such as number of blocks, connections, subsystems, and  linked blocks\footnote{\url{https://www.mathworks.com/help/simulink/ug/creating-and-working-with-linked-blocks.html}}) we implemented the \xAppName{}-Metrics tool~\cite{tool:SlnetMetrics} on top of Simulink's APIs. While our \SL{} installation and toolbox configuration~\cite{SLNET:WIKI} cannot compile a significant portion of \xAppName{} models (mostly due to missing toolbox licenses), these APIs still compute metrics for these non-compiling models, except for three metrics (algebraic loops, cyclomatic complexity, and compile time).

\xAppName{}-Metrics failed to compute metrics for \SLNETnoMetricModels{}/\SLNETlegalTotalModel{} 
models (21 from GitHub, 67 from MATLAB Central). Most of these \SLNETnoMetricModels{} were due to \SL{} version issues (missing \SL{} toolboxes, model name conflicts with a keyword or toolbox file name) and bugs introduced by manually-edited model files. \xAppName{} does not include metrics for these \SLNETnoMetricModels{} models and thus also ignores them for the above duplicate-via-bijection removal.

\xAppName{}-Metrics collects each model's hierarchical depth, solver type, simulation mode, target hardware, and use of S-functions and model references. While \xAppName{} models contain elements from the state-machine toolbox Stateflow, Stateflow is out of scope and our metrics do not count the Stateflow-contents of a \SL{} block.

Unlike \CorpusCurated{}, \xAppName{}-Metrics does not count
nested blocks imported from libraries or their connections (aka ``masked subsystems''). This mirrors procedural code metrics, which also do not count LOC a program imports from a library. 
As \CorpusCurated{}'s counting of such imported blocks approximates the model's overall conceptual complexity~\cite{tool:howmanyblocks}, Table~\ref{tab:replication} also includes these counts. As an example, the Figure~\ref{fig:toySimulinkModel} model imports blocks from the Simscape toolbox, yielding a \CorpusCurated{}-style block count of 907 with 919 connections.

The \SL{} API labels only 9 \xAppName{} models as a test harness, likely because many open-source projects do not have the required ''Simulink Test'' license to develop such tests. 
Beyond this official classification \xAppName{} contains likely ``work-around'' test harnesses. The \CorpusCurated{} metrics tool heuristically matches model and folder names with ``test'' and ``harness'' and \xAppName{} labels such models separately.

We performed sanity checks on the model metrics other papers reported about industrial models (block count, etc.). We also randomly sampled from the top 100 largest models in \xAppName{}. Based on the sampled models’ documentation we are confident that these were real human-created (non-synthetic) models.

\section{Potential Research Directions}

Since most industrial models are proprietary \xAppName{} is unlikely to reflect their distribution. Instead, the goal is to provide the largest possible redistributable self-contained corpus of non-synthetic models. Different research projects will require different \xAppName{} subsets (e.g., many small models for training deep-learning classifiers vs. large models to evaluate a technique’s scalability), which the SQL metadata database facilitates. Having more models is better, especially in deep learning, but also when trying to understand the breadth of modelling practices, or when looking for edge cases (e.g., to test model analysis tools). Following are example directions.

\balance 
While there has been significant interest in other software engineering areas~\cite{others:mlforse:systematicreview,others:mlforse:mlfordesignmatterdetect,others:MLforSE:codecompletion}, applying machine learning is relatively under-explored in model driven engineering~\cite{simulink:masterthesis:modelcompletion,CPS:MLforsecurity:Chen}. To work well, many machine learning and deep learning algorithms require large training sets. \xAppName{} with its many models and rich metadata is thus well-suited. For example, a \xAppName{} subset has been used to train a deep learning model for random \SL{} model generation, to find bugs in the Simulink toolchain~\cite{lab:SLGPT}. Due to their smaller size, this would have not been possible with the earlier corpora.

Due to the lack of easily available open-source models that fit certain characteristics,
recent work reverted to evaluating tools on synthetic models~\cite{Chowdhury20SLEMI}. \xAppName{} offers a complimentary (and often preferred) evaluation option with human-authored models.

Recent work including in clone detection, refactoring, model slicing, and model smells has relied on evaluations with few proprietary Simulink models~\cite{simulink:clone:Deissenboeck,simulink:refactor:pantelic,simulink:slicing:ReichertG12,simulink:smell:detectandHandleModelSmell,Boll:2020}. For example, Deissenboeck et al.~\cite{simulink:clone:Deissenboeck} evaluated their clone detection approach on a single proprietary Simulink model with 20k blocks. 
Complementing such evaluations with a variety of open-source models from \xAppName{} could make such studies more general and easier to replicate.

Understanding modeling practices would enable researchers to tune their tools to how engineers use \SL{} in various settings. For example, SLforge guides its random model generation by how often blocks appear in 391 open-source models~\cite{chowdhury2018icse}. The larger size of \xAppName{} could thus, e.g., yield useful insights for tool design.

There may also be interesting correlations between metrics, maybe connecting model metrics to project metrics (e.g., model size metrics with project engagement). More generally, \xAppName{} could contribute to a deeper understanding of model modularity, comprehension, quality, and maintainability~\cite{simulink:akasoftware_Dajsuren, simulink:Marta,simulink:metrics:comprehension_antonio2014set,simulink:metric:integration_testing)}.

While \xAppName{} is unlikely to exactly represent closed-source development, the precise shape of this relation is an open question. For example, for the related domain of Object
Constraint Language (OCL) expressions~\cite{ocl:definitiveguide:Cabot}, Mengerink et al. found the distribution of expression complexity mined from GitHub projects reflects the distribution in closed-source projects, so open-source projects can be used as a proxy for industrial projects~\cite{ocl:dataset:Noten,ocl:notsdifferentfromopensource:Mengerink}.

\section{Threats to validity}

Due to its search heuristics \xAppName{}-Miner may miss \SL{} models (e.g., by missing some of the non-documented RSS feed URLs). Furthermore, since \xAppName{} contains only redistributable projects, results may not be representative of all open source \SL{} projects. On the flip side, while removing forks and duplicates, \xAppName{} 
likely contains clones (from near-duplicate projects to adapted model portions), which can be an opportunity for clone-based research (and a challenge for others). Finally, \xAppName{}-Metrics calls the Check API of \SL{} R2019b. While this API has been available since \SL{} R2017b, its behavior may change across releases and thus yield different metric values in future Simulink versions.

\section{Conclusions}

\xAppName{} is the largest corpus of freely available third-party \SL{} models. 
\xAppName{} is 8~times larger than the largest previous \SL{} corpus, includes fine-grained metadata, is constructed automatically, is self-contained, and allows redistribution.

\begin{acks}
Christoph Csallner has a potential research conflict of interest due to a financial interest with Microsoft and The Trade Desk. A management plan has been created to preserve objectivity in research in accordance with UTA policy. This material is based upon work supported by the National Science Foundation (NSF) under Grant No. 1911017 and a gift from MathWorks.
\end{acks}

\bibliographystyle{ACM-Reference-Format}
\bibliography{ref}
\end{document}